\documentclass[aps,pre,floats,showpacs,superscriptaddress]{revtex4}
\usepackage{graphicx,epsfig}
\usepackage{times}
\usepackage{graphics,dcolumn,bm,epic,eepic,float}
\usepackage{amssymb,amsmath,multirow,rotate,color}
\begin{document}

\title{Phase synchronization on scale-free and random networks in the presence of  noise  }

\author
{Hamid Khoshbakht{\footnote {Electronic address:
hamidkhoshbakht@ph.iut.ac.ir}}}

\author{Farhad Shahbazi{\footnote {Electronic address:
shahbazi@cc.iut.ac.ir}}}

\author{Keivan Aghababaei Samani{\footnote {Electronic address:
samani@cc.iut.ac.ir}}}

\affiliation {\it Dept. of Physics , Isfahan University of
Technology, 84156-83111, Isfahan, Iran}

\date{\today}

\begin{abstract}
In this work we investigate the  stability of synchronized states
for the Kuramoto model on scale-free and random networks in the
presence of white noise forcing. We show that for a fixed coupling
constant, the robustness of the globally synchronized state
against the noise is dependent on the noise intensity on both
kinds of networks. At low noise intensities the random networks
are more robust against losing the coherency but upon increasing
the noise, at a specific noise strength the synchronization among
the population vanishes suddenly. In contrast, on  scale-free
networks the global synchronization  disappears continuously at a
much larger critical noise intensity  respect to the random
networks.
\end{abstract}

\pacs{
05.45.Xt, 
02.50.Ey, 
89.75.Fb. 
} 


\maketitle

\section{Introduction}
Collective behavior in a population of individuals is of great
importance in many areas of physics, biology, social sciences and
many other disciplines\cite{sync,bio,winfree}. One of the most
celebrated collective behaviors is the case of phase
synchronization among a population of interacting self-sustained
oscillators,  in which all members tend to oscillate coherently
with more or less the same phase. A population of coupled phase
oscillators with mutual sine interaction between the pairs, is
known as the {\it Kuramoto model}~\cite{kuramoto}. This model is
introduced by Kuramoto ~\cite{kuramoto1} and has extensively been
investigated by many authors (see \cite{acebron,boccaletti,book1}
and references therein).

Considering a network of coupled rotors (phase oscillators), many
factors such as coupling strength\cite{cs},  time delayed
interactions \cite{td}, individual frequency
distribution\cite{acebron,bimodal}, topology of network
\cite{topology} and noises\cite{acebron,noise}, affect the path
toward the full synchronization.

The synchronization of the deterministic  Kuramoto model with
random distribution of rotor frequencies and initial phases, has
been  studied  on the scale-free networks  by Moreno and Pacheco
\cite{SFsync}. There, it has been shown that the onset of
synchronization occurs through a continuous transition at a small
value of coupling with a critical exponent  near $0.5$. This
resembles the mean-field behaviour, except  that in the case of
scale-free networks the critical coupling at which the rotors
begin to get synchronized, is much smaller respect to all-to-all
networks. They have also managed to find that in the complete
synchronized  state, dependence of the recovery time respect to
the node degree is a power law function with exponent close to
-1, which shows the robustness of highly connected nodes (hubs)
against perturbations.

The comparison between  the synchronizability  of the Kuramoto
model on Erd\"{o}s-R\`{e}nyi(ER) and scale-free networks has been
carried out recently by Gom\'{e}z-Garde\~{n} \textit{et al}
\cite{gomez1,gomez2}. In these references, the authors have found
that while the onset of global synchronization occurs at smaller
value of coupling for the scale-free networks, but tendency
toward the global coherence grows suddenly to larger extends for
ER networks, at higher couplings. The reason for such behaviour
is that the giant cluster in the heterogeneous networks (such as
scale-free networks), originating  from a central core of high
connectivity (hub), grows continuously by attaching the smaller
clusters to it upon increasing the coupling constant. In
contrast, for the homogeneous networks, the evolution toward full
synchronization would be boosted up by merging many small
clusters spreading uniformly though out the networks, when the
coupling is large enough.

In this paper our focus is on the effect of white noise forcing
on the synchronization of Kuramoto model for two types of
networks:  scale-free networks introduced by Barab\'{a}si and
Albert(BA)~\cite{BA} and random networks\cite{ER}. The paper is
organized as follows: in section II, we introduce the stochastic
Kuramoto model and express briefly some results   for all-to-all
networks, section III, is devoted to the simulation results  and
the conclusion will be presented in section IV.

\section{Stochastic Kuramoto model on a complex network}

Consider  a system composed of $N$ rotors with the intrinsic
frequencies, denoted by $\omega_{i}$, on the top of a complex
network consisting of $N$ nodes. The stochastic Kuramoto model on
such a network is described by the following set of equations:
\begin{equation}
 \frac{d\theta_i}{dt}=\omega_{i}+\lambda\sum_j
 a_{ij}\sin(\theta_j-\theta_i)+\eta_i(t)\;, \hspace{0.5cm}i=1\cdot\cdot\cdot N,
 \label{kuramoto}
 \end{equation}
where $\theta_i$ is the phase of the rotor on the node $i$,
$\lambda$ is the coupling constant and $a_{ij}$ is an  element of
the connectivity matrix, which takes the value $a_{ij}=1$ if $i$
and $j$ are linked together and $0$ otherwise. $\eta_{i}$ is the
random force applying on the $i$-th rotor, and usually is chosen
as a white noise with zero mean-value. The spatial-temporal
correlation of such a noise is given by:

\begin{equation}
\langle\eta_i(t)\eta_j(t')\rangle=2D\delta(t-t')\delta_{ij},
\end{equation}
 where $D$ is  variance  of
 the noise.

The stochastic Kuramoto model on an all-to-all network has been
investigated analytically by Acebr\'{o}n \textit{et al}
\cite{acebron}, who have  shown that  taking a  one peaked
symmetrical  frequency distribution $f(\omega)=f(-\omega)$ for
oscillators, there would be  a critical coupling constant
$\lambda_c=2/[\pi f(0)]$ above which the network begins to get
synchronized. Near this critical point, the order parameter obeys
a power law relation, namely
    \begin{equation}
    r\sim
    \sqrt{\frac{-16(\lambda-\lambda_c)}{\pi\lambda_c^4f''(0)}}.
    \label{r-scale}
    \end{equation}
They have also shown that for a Lorentzian frequency distribution
$f(\omega)=(\gamma/\pi)/(\omega^2+\gamma^2)$, the incoherent
solution  is linearly stable for points $(\lambda,D)$ above the
critical line $D=-\gamma+\lambda/2$. In terms of coupling
strength, this is also linearly stable for
$\lambda<\lambda_c=2D+2\gamma$.

In the next section we numerically integrate Eq.(\ref{kuramoto})
on scale-free and random networks and compare the results.

\section{simulation results for scale-free and random networks}

To create a  scale-free network with average connectivity $\langle
k \rangle=2m$, we use the BA algorithm \cite{BA}. In this
procedure, starting from $m_{0}$ initial nodes all connected to
each other, at each step one attaches a newly entering  node to
$m\leq m_{0}$ elder ones such that the nodes with higher
connectivity  have larger probability (proportional to their
degree) to get connected with this new one. Repeating this stages
provides us with a network whose degree distribution obeys a
power law function as $P(k)\sim k^{-\gamma}$ with $\gamma=3$. For
producing ER network composed of $N$ nodes and with the same
average degree per node($\langle k \rangle=2m$), it is enough to
distribute $Nm$ edges between randomly chosen pair of
nodes\cite{ER}.

In this work, we set $m=10$ and select a  delta function
distribution for intrinsic frequencies,
$f(\omega)=\delta(\omega-\omega_{0})$. Changing the reference
frame to a rotating one with rotation frequency $\omega_{0}$,
enable us to set $\omega_{i}=0$ for all rotors in
Eq.(\ref{kuramoto}).  We also pick $\eta_{i}(t)$ out of a box
distribution in the interval $-g/2<\eta<g/2$, hence its variance
is given by $D=\frac{g^2}{24}$. Using Ito's formalism for
integration of a stochastic function\cite{ito}, one obtains the
following discrete equation from Eq.(\ref{kuramoto}):

 \begin{equation}
\theta_i(t+dt)=\theta_i(t)+\lambda\left[\sum_ja_{ij}\sin(\theta_j(t)-\theta_i(t))\right]dt+\eta_i(t)\sqrt{dt}+O(dt^2),
 \end{equation}
  where in the Ito's picture, $\eta_{i}(t)$ is evaluated at the
initial point of the time interval $[t, t+dt]$. Time step, $dt$,
is taken small enough to reduce the computational error. The
initial values of $\theta_i$ are randomly drawn from a uniform
distribution in the interval $[-\pi,\pi]$. To characterize the
global phase coherency, we define the following order parameter:
 \begin{equation}
 r(t)=\langle|\frac{1}{N}\sum_{j=1}^{N}e^{i\theta_i(t)}|\rangle,
 \end{equation}
 which $\langle\cdot\cdot\cdot \rangle$ means the averaging over
 different realizations of noise and initial conditions. In the stationary regime the time
 argument of $r(t)$ can be omitted and one can replace the
 averaging over realizations by time averaging. The order parameter takes the value $0\leq r \leq 1$, where $r=0$ corresponds to the disordered
 phase  and $r=1$ characterizes  the full synchronized state.

  In the absence of noise for  scale-free network, we found that the rotors get synchronized for
 a very small value of coupling around $\lambda=0.03$.  We
 choose a large enough value for the coupling to make sure that the
 system is in the full synchronized state when the noise is
 turned off, then increase  the noise intensity until the global coherency
 vanishes at the critical value of the noise strength, $g_{c}$.

 In addition to the order parameter $r$, for better specifying  the
 transition from coherency to decoherency, we introduce the
 Binder's forth cumulant which is defined as:

 \begin{equation}
 u=1-\frac{\langle r^4\rangle}{3\langle r^2\rangle^2}.
 \label{binder}
 \end{equation}
It is easy to see that in coherent phase, where $\langle r
\rangle$ is nonzero, $u$ takes the value $2/3$ in the large $N$
limit, while upon vanishing the global synchronization($\langle r
\rangle$=0) this quantity falls down to $1/3$ in thermodynamic
limit. The much smaller numerical errors in computation of the
Binder's forth cumulant rather than the order parameter, makes it
more advantageous for determination of the position and treatment
of the coherency-decoherency phase transition.

\begin{figure}[h]
\epsfig{file=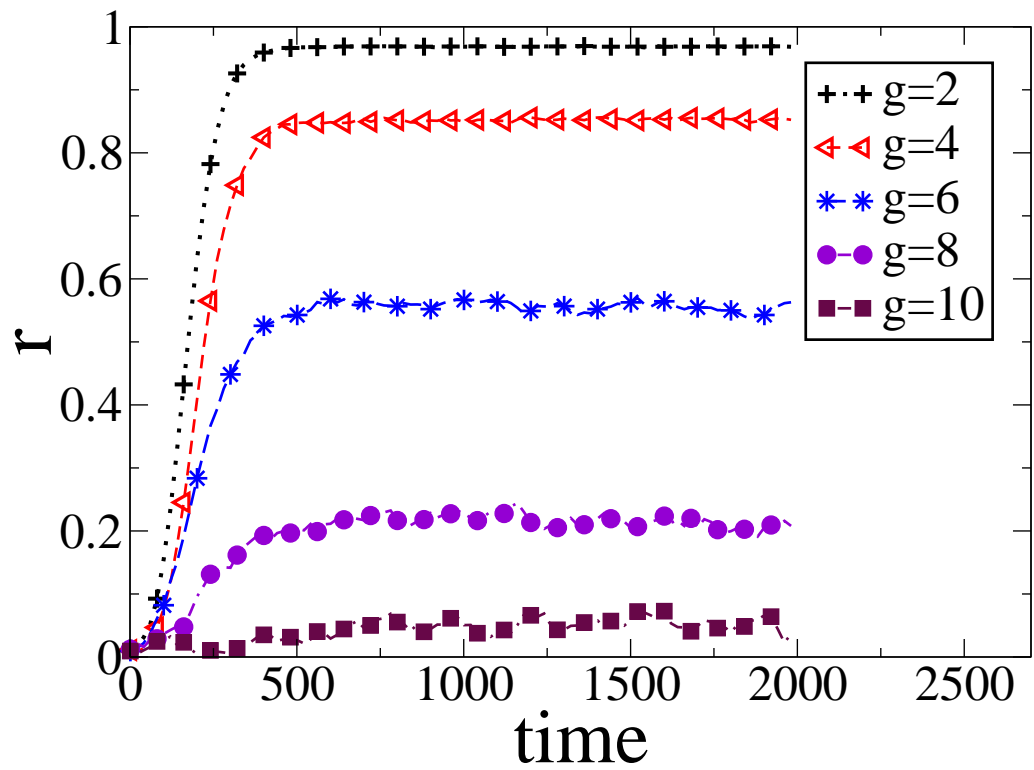,width=4.5in,angle=0,clip=1} \caption{Order
parameter versus time for different noise intensities on the
scale-free network. The results are obtained for  coupling
constant $\lambda=0.2$ and the number of nodes is $N=10^4$.}
\label{fig1}
\end{figure}

\begin{figure}[h]
\epsfig{file=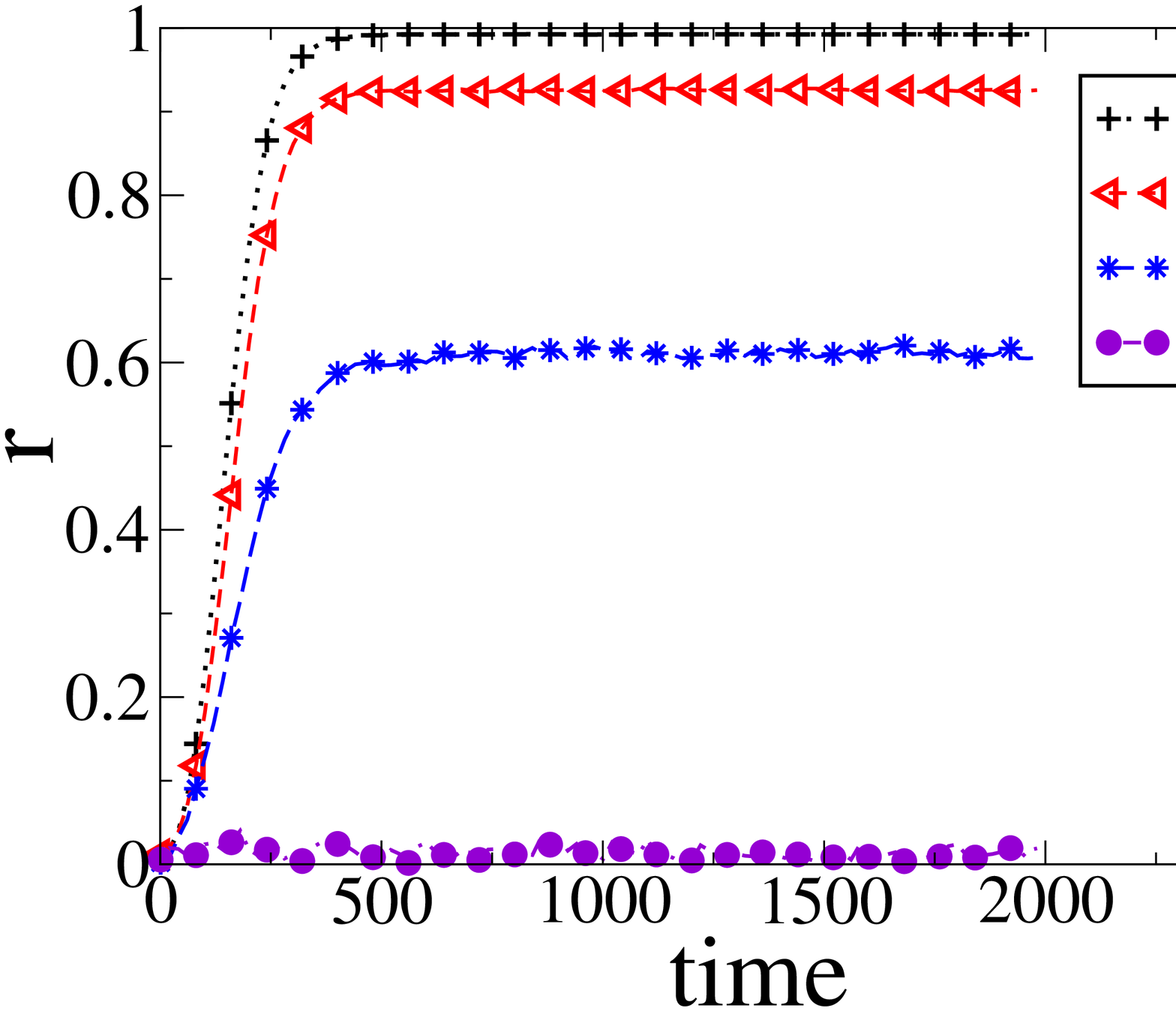,width=4.5in,angle=0,clip=1} \caption{Order
parameter versus time for different noise intensities on the
random network. The results are obtained for coupling constant
$\lambda=0.2$ and the number of nodes is $N=10^4$.} \label{fig2}
\end{figure}

 Figs.\ref{fig1} and \ref{fig2} represent the time
 dependence of order parameter for scale-free and random networks composed of $N=10^4$ rotors,
 respectively. To derive these data, we put $dt=0.01$, $\lambda=0.2$  and
 averagings have carried out over 100 different realizations of
 noise and initial phase configurations, for the noise intensities increasing  from $g=2.0$ by step $\Delta g=2.0$. From these figures, one
 finds that after about 500 time steps the system reaches the stationary for all noise
 intensities, and  obviously the global coherency vanishes at
 larger coupling values for the scale-free network (around $g=10.0$)
 respect to the random network (around $g=8.0$).

 In what follows, to find the
 dependence of the order parameter as well as
 Binder's forth cumulant on the noise intensity, we fix the number of nodes to  $N=10^4$, and the averagings are carried on  $10^4$ time
 steps after skipping $2000$ initial steps, where the system is surely in the stationary state.

\begin{figure}[h]
\epsfig{file=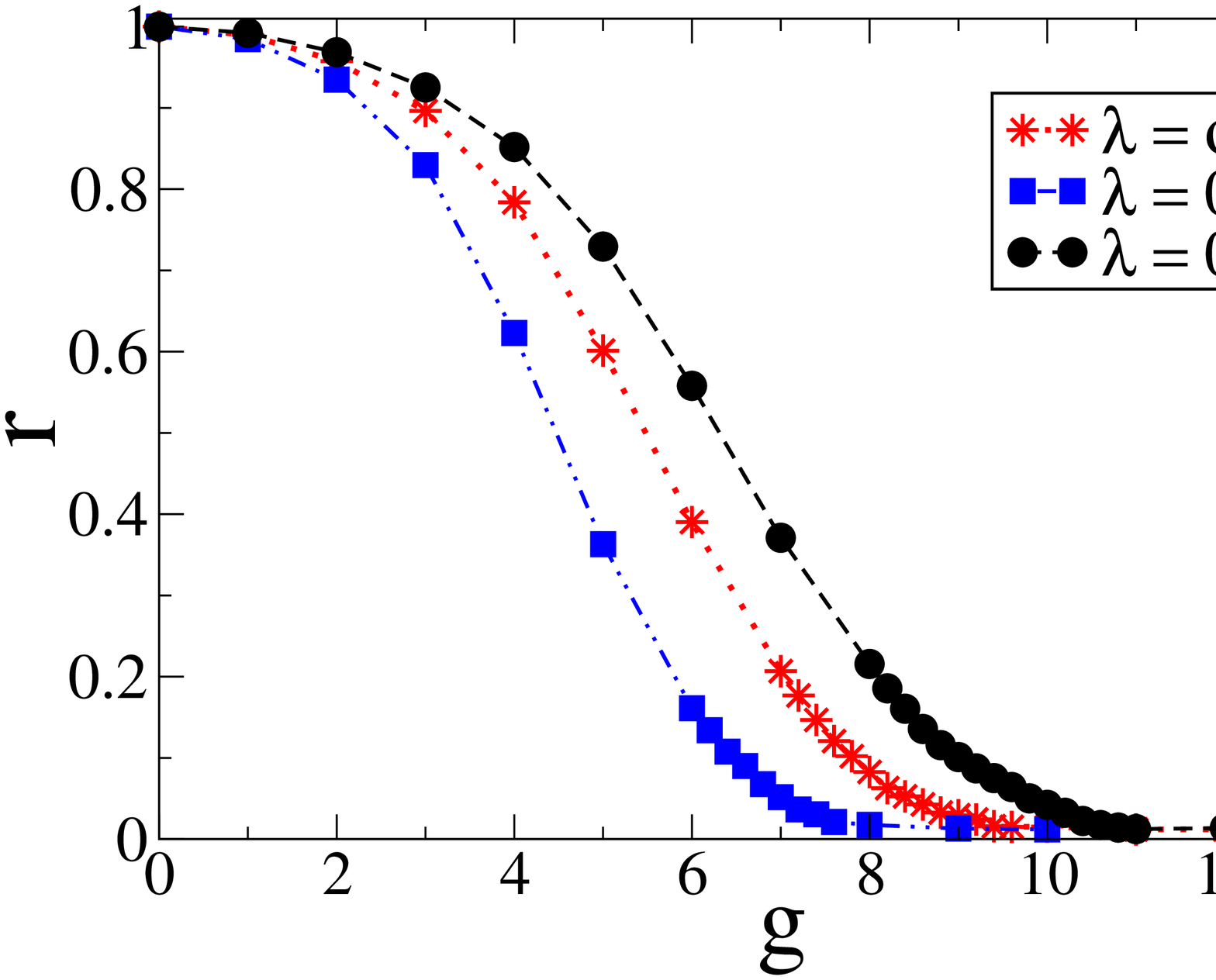,width=4.5in,angle=0,clip=1} \caption{The
order parameter versus the noise intensity for the scale-free
network. The results are obtained for three coupling values
$\lambda=0.1,0.15,0.2$ and $N=10^4$ phase oscillators.}
\label{fig3}
\end{figure}

\begin{figure}[h]
\epsfig{file=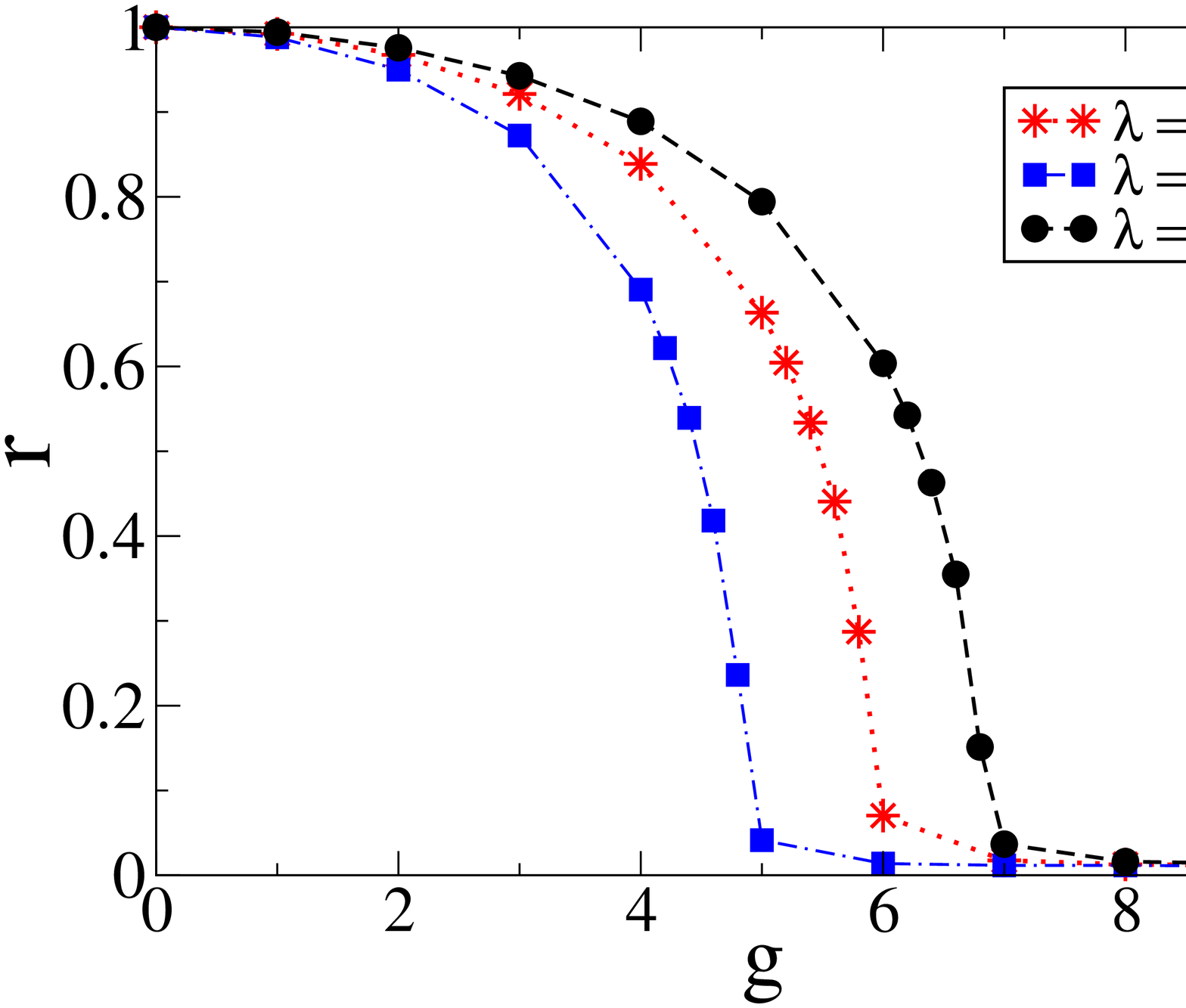,width=4.5in,angle=0,clip=1} \caption{The
order parameter versus the noise intensity for the random
network. The results are obtained for three coupling values
$\lambda=0.1,0.15,0.2$ and $N=10^4$ phase oscillators.}
\label{fig4}
\end{figure}

\begin{figure}[h]
\epsfig{file=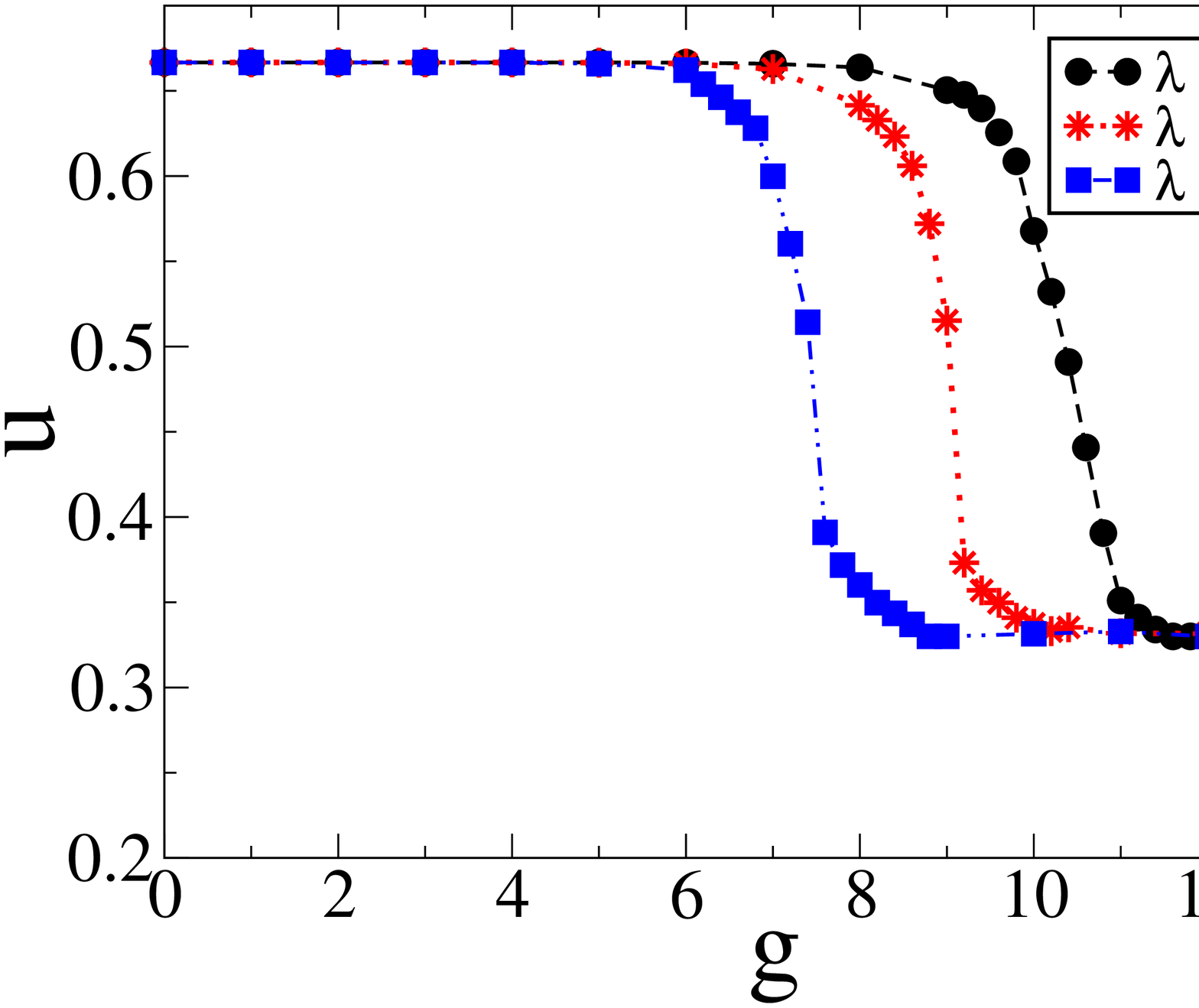,width=4.5in,angle=0,clip=1} \caption{The
Binder's forth cumulant (Eq.\ref{binder}) versus the noise
intensity for the scale-free network. The results are obtained
for three coupling values $\lambda=0.1,0.15,0.2$ and $N=10^4$
phase oscillators.} \label{fig5}
\end{figure}

\begin{figure}[h]
\epsfig{file=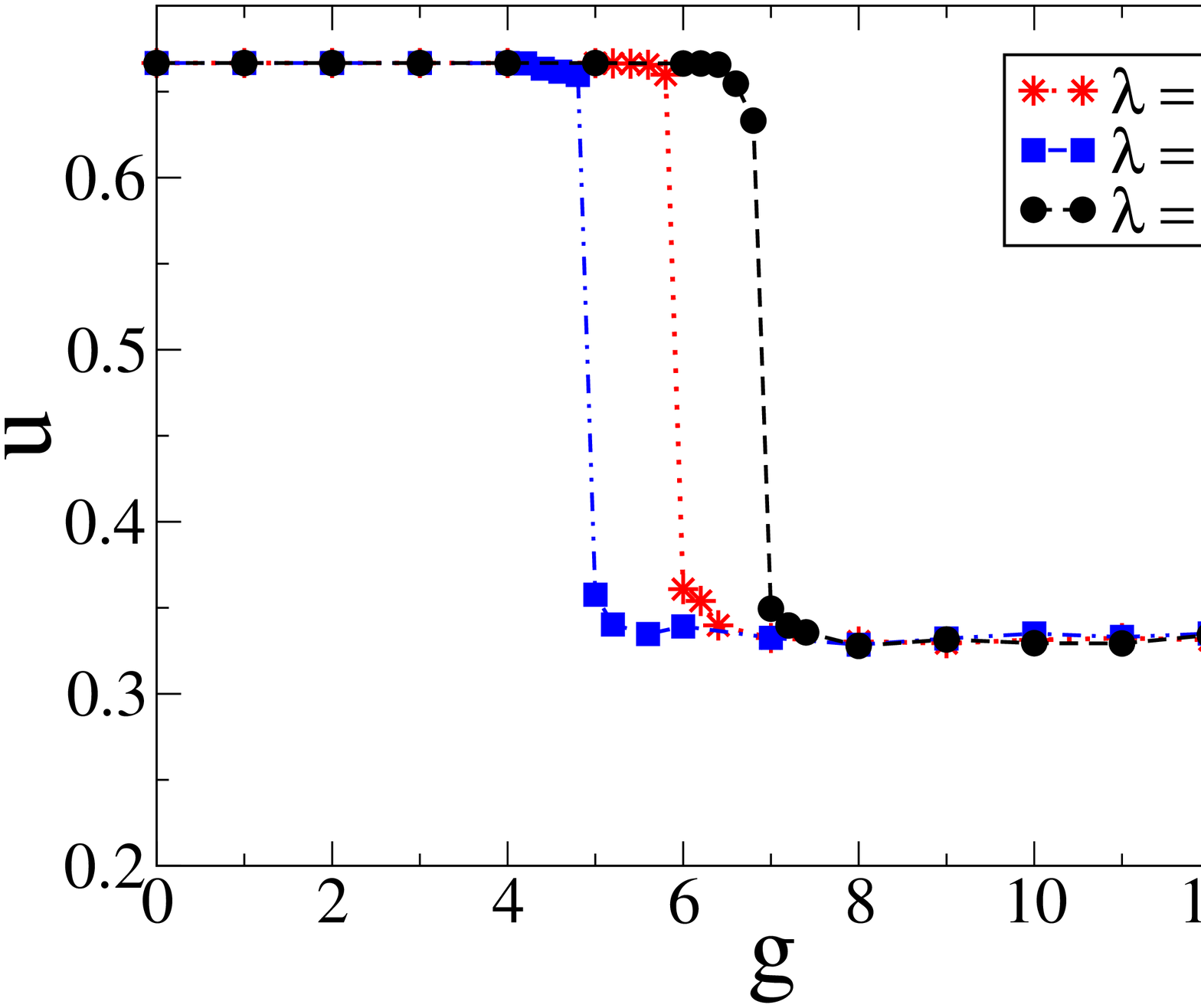,width=4.5in,angle=0,clip=-2} \caption{The
Binder's forth cumulant (Eq.\ref{binder}) versus the noise
intensity for the random network. The results are obtained for
three coupling values $\lambda=0.1,0.15,0.2$ and $N=10^4$ phase
oscillators.} \label{fig6}
\end{figure}

 In Figs.\ref{fig3} and \ref{fig4} we have depicted  the order
 parameter, $r$, versus the noise intensity, $g$, for scale-free and random networks, for three coupling constants $\lambda=0.1,0.15,0.2$.
 Similar graphs for the Binder's forth cumulant, $u$, is represented in
 Figs.\ref{fig5} and \ref{fig6}.

  For comparison,  the noise intensity variations of $r$ and $u$ have been depicted for both scale-free and random networks
  in Figs.7a and 7b.

   By inspecting these figures one can extract two essential results:\\
 (i) Synchronizability  of each kind of networks depends on the
    coupling strength, such that at small noise intensities the order parameter for random
    network falls more slowly than scale-free's, so it is more robust than SF network against the noise while at large noise
    intensities  the situation is vice versa (see Fig.7a). The critical noise intensity ($g_{c}$)
    at which the transition from synchronized to un-synchronized state occurs is larger for the  scale-free network. So the coherent state in
    SF network persists  more against the  noise than the random network with the same average degree and coupling constant.\\
 (ii) The  coherency among the population of rotors  destroys  smoothly by increasing the  noise intensity  in the scale-free
 networks, while in the random networks the synchronization disappears by a sudden fall at the transition point. These behaviours are more
 apparent from the treatment of Binder's forth cumulant shown in Figs.\ref{fig5}, \ref{fig6} and
7b. Then the order-disorder transition in SF networks resembles
the continuous transitions  in equilibrium critical phenomena,
while
 the transition in random networks is discontinuous-like.

 Referring to Gom\'{e}z-Garde\~{n}es \textit{et
 al}\cite{gomez1,gomez2}, a nice  explanation of our results are as follows. In homogeneous systems such as random
 networks, starting from the fully synchronized phase, when we turn on the noise, some incoherent clusters  with more or less the same size begin
 to form. At low noise intensities, the size of these clusters
 are small and they are well separated, but by increasing the
 noise they get larger and connected to each other at intermediate noise intensities. At this point,
  the locally synchronized regions are not coherent anymore, so a big
 drop occurs for the order parameter. This is much like the first
 order phase transitions in equilibrium statistical mechanics,
 where the ordered and disordered phases coexist at the transition point. On the other
 hand, the fully coherent state in the SF networks is founded
 around a core consists of  nodes with high connectivity
 (hubs). When noise is applied on such state, the un-synchronized
 parts leave this giant cluster one by one, leading to continuous
 destruction of global coherency.

\section{conclusion}

In summary, we numerically investigated the stability of the
global phase synchronized state in  Kuramoto model on the top of
 scale-free and random networks, under white noise forcing on
each oscillator. Our results  emphasize on the fact that the
 stability of the synchronized phase is dependent on the noise
 strength, such that at low noise intensities the random networks
 are more stable against loosing the coherency, while at intermediate noise intensities, the coherency falls abruptly in such networks. However, in
 scale-free networks the  coherency among the rotors  decreases  smoothly and also persists up to  larger extends  of  noise intensity.
 Therefore, our findings  confirm the picture presented by
Gom\'{e}z-Garde\~{n}es \textit{et
 al}\cite{gomez1,gomez2}, that in heterogeneous networks the  giant cluster
 formed around a core of hubs, grows (falls) continuously by increasing (decreasing) the coupling
 or by lowering (rising) the noise intensity. On the contrary, in
 homogeneous systems such as  random networks, by increasing the noise intensity, the coalescence of
 un-synchronized clusters  which are uniformly distributed over the network,
 results in a sudden fall in the global synchronization  at the transition point. So the more complex is a system,
 more predictable it is.

 This work sheds more light on the different aspects  of nonlinear dynamics on the top of homogeneous
 and heterogeneous network topologies and we hope that it promotes
 more researches on this very interesting problem.

\newpage

\newpage

\newpage

\newpage

\newpage

\newpage

\newpage
\begin{figure}[t]
\epsfig{file=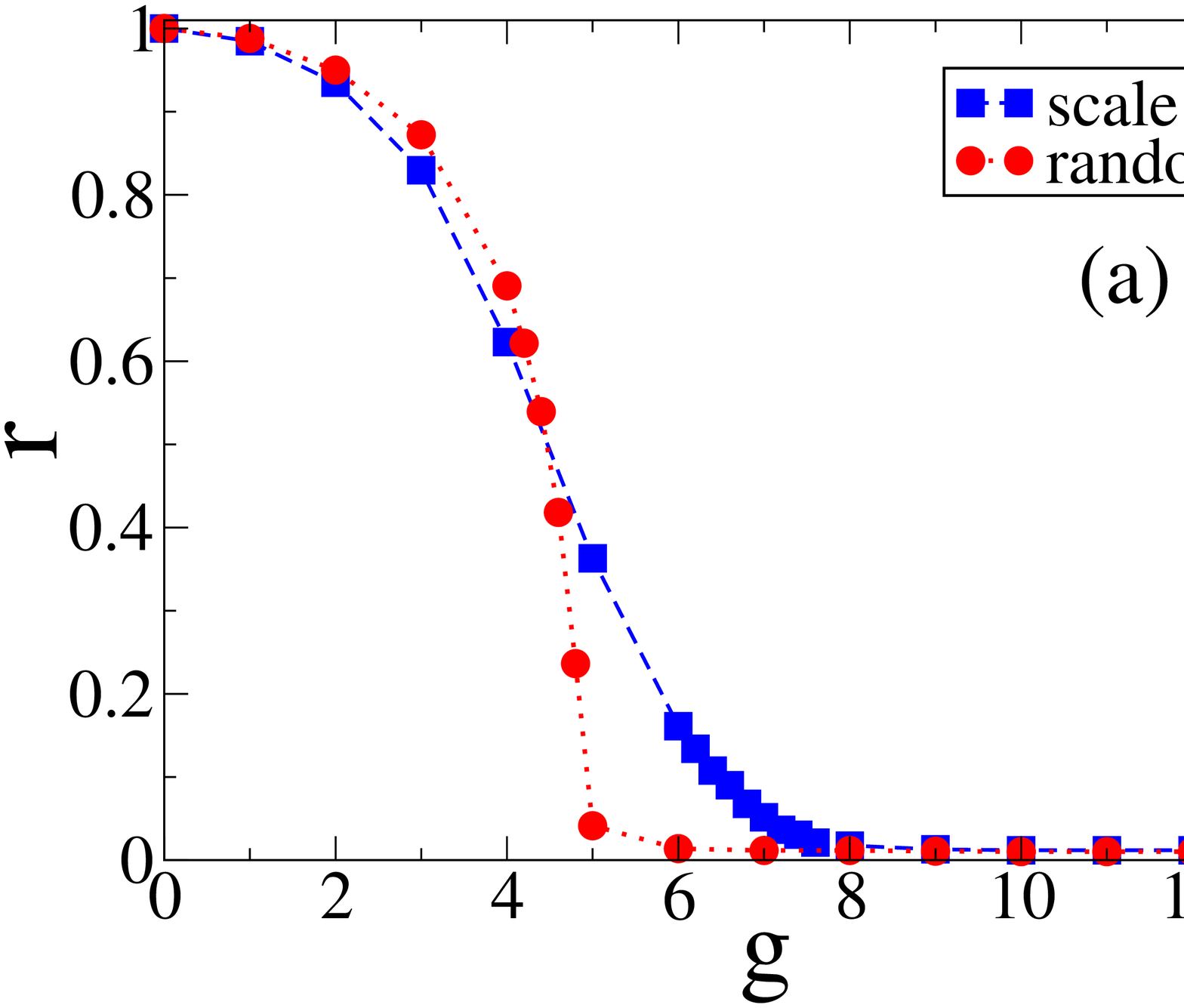,width=4.in,angle=0,clip=1}\nonumber
\end{figure}
\begin{figure}[b]
\epsfig{file=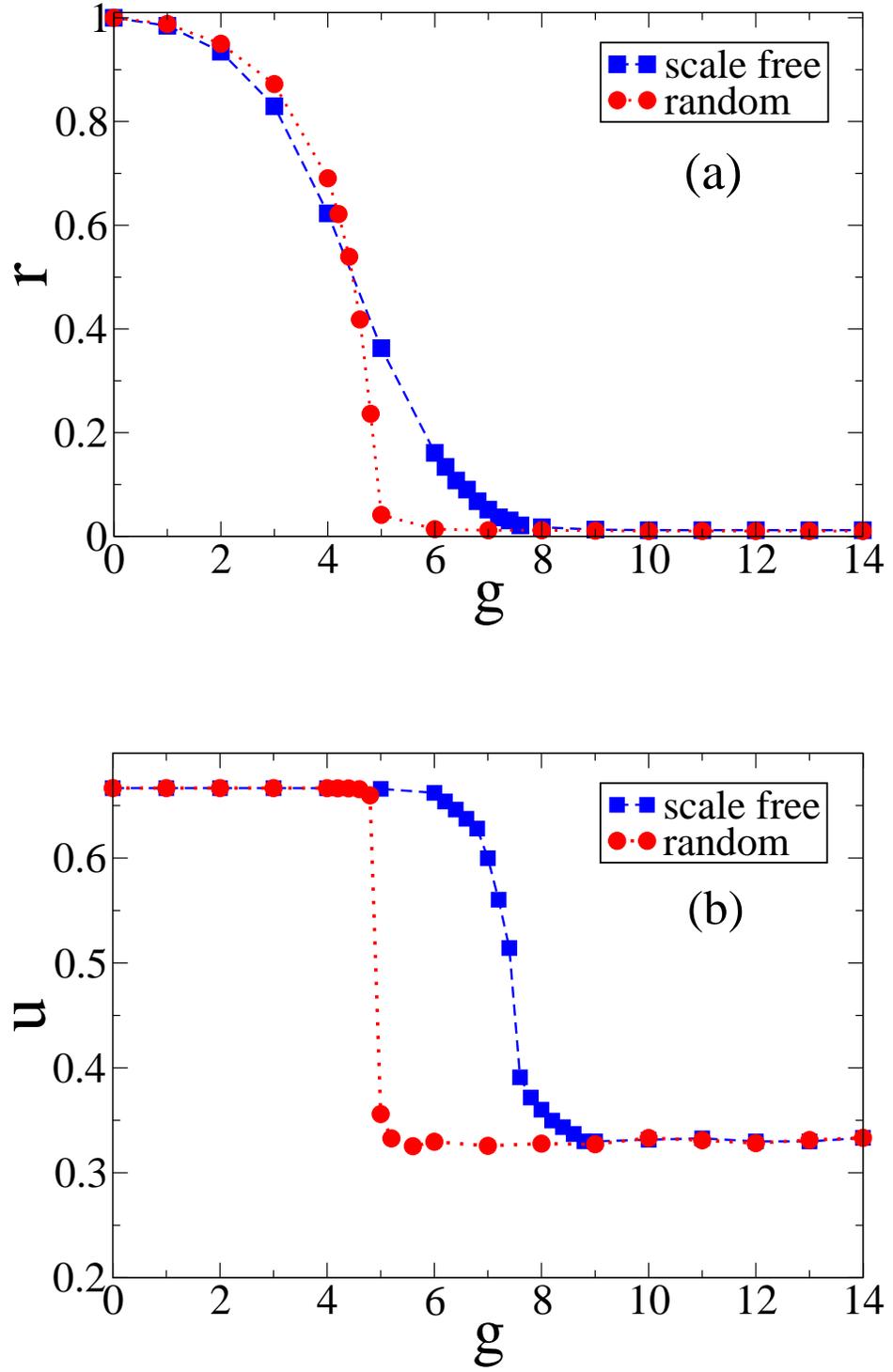,width=4.in,angle=0,clip=2} \caption{Noise
intensity dependence of (a) Order parameter and (b) Binder's
forth cumulant  for the
    scale free and random network at $\lambda=0.1$.} \label{fig7}
\end{figure}

\end{document}